\documentclass[aps,prl,epsfigure,twocolumn,superscriptaddress]{revtex4-2}
\usepackage[colorlinks=true,linkcolor=blue,urlcolor=blue,citecolor=blue,pdfusetitle]{hyperref}
\usepackage{amsfonts}
\usepackage{graphicx} 
\usepackage{color}
\usepackage{amsmath}
\usepackage{kantlipsum}
\usepackage{soul}
\usepackage{bibunits}
\usepackage{physics}
\usepackage[toc,page]{appendix} 

\usepackage[normalem]{ulem}

\usepackage{float}

\begin{document}

\title{Strong Collective Chiroptical Response from Electric-Dipole Interactions in Atomic Systems}

\date{\today}

\author{Marcella L. Xavier}
\email{marcella\_lx@hotmail.com}
\affiliation{Departamento de F\'{\i}sica, Universidade Federal de S\~{a}o Carlos, Rodovia Washington Lu\'{\i}s, km 235 - SP-310, 13565-905 S\~{a}o Carlos, SP, Brazil}

\author{ Felipe A. Pinheiro}
\email{fpinheiro@if.ufrj.br}
\affiliation{Instituto de F\'{\i}sica, Universidade Federal Rio de Janeiro, 21941-972 Rio de Janeiro, RJ, Brazil}

\author{Romain Bachelard}
\email{romain@ufscar.br}
\affiliation{Departamento de F\'{\i}sica, Universidade Federal de S\~{a}o Carlos, Rodovia Washington Lu\'{\i}s, km 235 - SP-310, 13565-905 S\~{a}o Carlos, SP, Brazil}

\begin{abstract}
Chiroptical responses in atomic systems are usually weak, as they arise from the interference between electric- and much weaker magnetic-dipole transitions. We show that atoms arranged in chiral geometries can instead exhibit a strong collective chiroptical response mediated entirely by electric-dipole interactions. Using a coupled-dipole framework, we identify a regime of pronounced chiroptical response emerging at subwavelength interatomic separations, which can be tuned by the probe frequency. This enhancement is directly linked to the formation of subradiant collective modes. Our results establish a fundamental connection between geometric chirality and collective light–matter interactions, opening new pathways for engineering and exploiting chiral optical responses in atomic systems.

\end{abstract}

\maketitle



{\em Introduction.} Chirality pervades nature across different length scales, from macroscopic objects to nanoscopic structures~\cite{wagniere2007chirality}. First introduced by Lord Kelvin, a chiral system is defined as one that cannot be superimposed onto its mirror image through any combination of rotations and translations~\cite{wagniere2007chirality}. Distinguishing between the two enantiomeric forms is essential in chemistry, molecular biology, and pharmaceutics, where molecular handedness critically determines the functionality of many compounds~\cite{maier2001separation}.
A wide range of strategies has been developed to separate and identify chiral molecules and particles, most prominently chemical methods~\cite{gubitz2001chiral}. However, these approaches are typically species-specific, often invasive, and rely on ensemble-averaged measurements. Probing chirality at the single-particle or single-molecule level remains a scientific and technological challenge due to the intrinsically weak nature of chiroptical signals~\cite{kumar2019recent,solomon2020nanophotonic}.

To address these limitations, plasmonic nanostructures have emerged as a powerful platform for enhancing chiroptical interactions by leveraging localized surface plasmon resonances~\cite{kim2022enantioselective,liu2020enantiomeric}, in which intense electric multipoles can be excited~\cite{im2024perspectives}. Beyond plasmonics, all-dielectric, high-index nanostructures and metamaterials have attracted growing interest, as their engineered magnetic multipolar resonances enable strong interference with electric modes, providing an alternative route to enhanced chiroptical responses~\cite{im2024perspectives,ho2017enhancing,garcia2019enhanced}.

Although chiroptical enhancement in single molecules and nanoparticles is traditionally achieved through the interference of electric and magnetic multipoles, purely electric strategies are emerging as a versatile and increasingly relevant alternative in photonics~\cite{D2CP01009G,Habibovi__2024}. Beyond photonics, atomic systems are particularly well suited for realizing controlled all-electric chiral excitations owing to their resonant nature and extremely narrow linewidths, which enable precise tuning of the chiroptical response~\cite{Muldoon_2012,PhysRevA.102.013306,NC_SingleOpticalTrap,PRXQuantum.6.020337}. In addition to chiroptical enhancement, atomic chiral platforms enable nonreciprocal directional emission~\cite{doi:10.1021/jp500319x,tilting}, chiral quantum optics~\cite{suarez2025chiral}, and chirality-dependent photon transport~\cite{PhysRevResearch.6.023200}. These capabilities open avenues for applications ranging from polarization control and optical logic~\cite{doi:10.1126/sciadv.abq8246} to quantum information processing~\cite{Lodahl_2017}. Moreover, atomic systems offer direct control over collective scattering phenomena, including superradiant emission~\cite{PhysRevResearch.6.023200}, further enhancing their appeal for tailoring chiral light–matter interactions. 

In this work, we demonstrate that small atomic assemblies arranged in chiral geometries can exhibit pronounced chiral optical responses that are tunable over frequency variations on the order of the linewidth. Remarkably, we find that the handedness of the response can be reversed simply by adjusting the probe frequency. The resulting strong circular dichroism (CD), enabling polarization-selective transmission, originates from cooperative scattering and the associated subradiant dynamics of the system. Moreover, our analysis of the wavelength dependence shows that, for small number of atoms within the electric-dipole approximation, the contrast between right- and left-circularly polarized transmitted light is maximized in the subwavelength regime.

{\em Chiroptical response induced by light-induced electric interactions.} We consider a set of $N$ two-level cold atoms at positions $\mathbf{r}_j$, driven by a weak coherent laser with Rabi frequency $\Omega$ and wavevector $\mathbf{k}$. Each atom is treated in the electric dipole approximation, with transition frequency $\omega_{0}$, and thus does not present chiral property itself. Under the rotating-wave and Markov approximations, the scattering dynamics is given by a set of $3N$ coupled dipole equations~\cite{Lehmberg1970RadiationFA,PhysRevA.81.053821,PhysRevA.94.023612,Araújo25062018}:
\begin{eqnarray}
    \dot{\beta}_{j}^{\zeta}(t) = \Bigg( i\Delta - &&\frac{\Gamma}{2} \Bigg)\beta_{j}^{\zeta}(t) -\frac{i\Omega_{\zeta}}{2}e^{i\textbf{k}\cdot\textbf{r}_{j}} \nonumber \\ &&- \frac{\Gamma}{2}\sum_{j\neq m}\sum_{\eta}G_{\zeta,\eta}(\textbf{r}_{j}-\textbf{r}_{m})\beta_{m}^{\eta}(t) \label{betas}
\end{eqnarray}
where $\beta_{j}^{\zeta}$ stands for the (complex) dipole components of atom $j$, with $\zeta \in \{-1, 0, +1\}$ referring to the polarizations $\textbf{\^{e}}_{\pm1} = 1/\sqrt{2}(\mp\textbf{\^{x}} - i\textbf{\^{y}}) $ and $\textbf{\^{e}}_{0} = \textbf{\^{z}}$. Here, $\Delta$ is the detuning between the laser frequency and the atomic transition and $\Gamma = d_{0}^{2}k^{3}/3\hbar\pi\epsilon_{0}$ is the single-atom decay rate, with $d_{0}$ the transition dipole moment and $\epsilon_{0}$ the free space permittivity. The relation between the Rabi frequency and the incident laser is given by $\Omega_{\zeta} = d_{0}\hat{\mathbf{e}}^{*}_{\zeta}\cdot\mathbf{E}_{L}/\hbar$.

The first right-hand term in Eq. \eqref{betas} corresponds to the single-atom dynamics and the second is the driving by the external electric field, while the third one stands for the light-induced coupling between the dipoles. This set of $3N$ equations rules the evolution of the amplitudes of the induced atomic dipoles in the laser frame, and their mutual coupling is encoded in the 
Green's dyadic function, $G_{\zeta,\eta}(\textbf{r})$~\cite{pinheiro2008statistics,fofanov2011dispersion}:
\begin{eqnarray}
    \overleftrightarrow{G}(\textbf{r}) = \frac{3}{2}\frac{\exp(ikr)}{ikr}&&\Bigg[\Bigg(1 + \frac{i}{kr} - \frac{1}{(kr)^{2}} \Bigg)\overleftrightarrow{\mbox{I}} \\ \nonumber&&+\Bigg(-1 -\frac{3i}{kr} + \frac{3}{(kr)^{2}}\Bigg)\frac{\textbf{r}\otimes \textbf{r}}{r^{2}}\Bigg]. \label{green_dyadic}
\end{eqnarray}
Note that this kernel corresponds to a $J = 0$ to $J = 1$ atomic transition. Furthermore, in the single-photon (linear optics) limit considered here, the dipole dynamics~\eqref{betas} can be derived through a fully classical approach, treating the particles as $N$ classical harmonic oscillators~\cite{PhysRevA.81.053821, Cottier2018}. Going beyond this regime to explore quantum chiral effects \cite{doi:10.1126/science.283.5403.814,Lodahl_2017,PhysRevA.109.043525,PRXQuantum.6.020101} will require a quantum description of the atomic operators and electric field~\cite{Lehmberg1970RadiationFA, Fayard2021, Pedersen2023, RubiesBigorda2023, Deepak2025}. In this work, numerical simulations were performed using the Julia package available at Ref. \cite{noel_disser,noel_code}.

\begin{widetext}
\begin{figure*}[t]
\centering
\includegraphics[width=18cm,height=4cm]{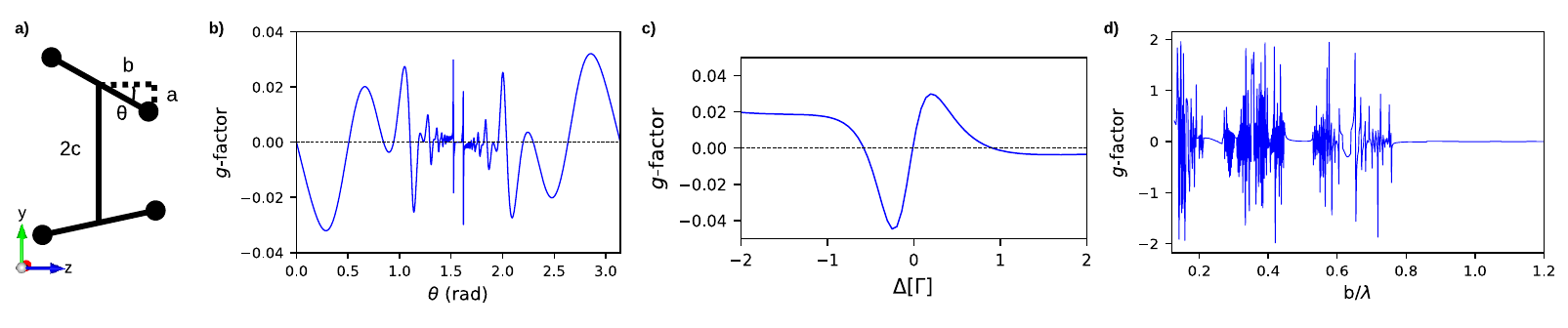}
\caption{a) Twisted H structure: Minimal intrinsically chiral system, with four non-coplanar atoms (electric dipoles) at positions $\mathbf{r}_{1} = (a,-c,b)$, $\mathbf{r}_{2} = (a,c,-b)$, $\mathbf{r}_{3} = (-a,c,b)$ e $\mathbf{r}_{4} = (-a,-c,-b)$, with $\tan\theta = a/b$. b) For different values of the angle one obtains distinct configurations of the structure, as evidenced by the dissymmetry factor (simulations realized for 
$b = 2.24/k$ and $c = 2.52/k$). c) Evolution of the CD as a function of the detuning $\Delta$; simulations realized for $b = 2.24/k$, $c = 2.52/k$ and $\theta = \pi/6$. d) The optical response is strongest in the subwavelengths, as shown by varying the system size. The relations between the lengths are here given by $c = 1.125 b$ and $a = b\tan{\theta}$, with $\theta = \pi/6$. }
\label{Figura_H}
\end{figure*}
\end{widetext}

{\em Twisted H configuration.} Let us first consider the simplest chiral system composed of $N=4$ atoms, hereafter referred to as twisted H structure, depicted in Fig.~\ref{Figura_H}(a). In order to break mirror symmetry,
the parameters $a$, $b$ and $c$ must be different and non-vanishing, with dipoles arranged in the extremities of
the edges~\cite{RevModPhys.71.1745}; the angle $\theta = \arctan(a/b)$ thus measures the deviation from the planar structure.

In the steady state, geometrical chiral properties show up in the polarization-dependent transmission and reflection coefficients, encoded in the definition of the $g$-factor that characterizes CD~\cite{cerdan25, https://doi.org/10.1002/chem.201501178}:
\begin{equation}
    g = 2\frac{T_{\mbox{\tiny RCP}} - T_{\mbox{\tiny LCP}}}{2 - (T_{\mbox{\tiny RCP}} + T_{\mbox{\tiny LCP}})} \label{eq_g_factor}
\end{equation}
where $T_{\mbox{\tiny RCP}}$ and $T_{\mbox{\tiny LCP}}$ are the transmission coefficients for the RCP and LCP light, respectively. 
The system is driven with a laser either RCP or LCP, and the corresponding transmission is computed in the forward direction as the ratio between the squared sum of the outgoing electric field (given by the sum of the incident one and that in the RCP or LCP polarization channel) and the intensity of the incident laser. Following Ref.~\cite{cipris:tel-03706978}, the scattered field can be computed from the third term of Eq.~\eqref{betas}, yielding 
\begin{equation}
\begin{aligned}
    E_\textrm{sc}^{\zeta}(\textbf{R},t) = -i\frac{dk_{0}^{3}}{6\pi\epsilon_{0}}\sum_{j}\sum_{\eta}G_{\zeta,\eta}(\textbf{R}-\textbf{r}_{j})\beta_{j}^{\eta}(t).
\end{aligned}
\end{equation}
As a Gaussian beam with waist $w_{0}=2\lambda$ is considered, the incident field $\mathbf{E}_{L}\exp(ikz)$ is described by

\begin{equation}
\begin{aligned}
    \mathbf{E}_{L}(\mathbf{R}) = E_0&\frac{w_0}{w(z)}\exp\left(-\frac{x^2+y^2}{w(z)^2}\right)\times\\ &\exp \left[i\left(k_L\frac{x^2+y^2}{2R(z)}-\arctan\left(\frac{z}{z_R}\right)\right)\right]\hat{\mathbf{e}}_{L}. 
\end{aligned}
\end{equation}

\noindent In this equation, $z_{R}$ represents the Rayleigh range, $R(z) = z[1+(z_{R}/z)^{2}]$ is the radius of curvature of wavefronts and $\hat{\mathbf{e}}_{L}$ denotes the field's polarization. 
The factor $g\in [-2; 2]$, with $|g|$ = 2 being associated to a maximal chirality, when the system is transparent to one polarization and backscatters efficiently the opposite-handed one~\cite{2020LSA.....9..139M}, with the sign of the dissymmetry factor accounting for the handedness of the system.

In Fig.~\ref{Figura_H}(b), we show the evolution of the $g$-factor with the angle $\theta$ of the twisted H structure. For $\theta = n \pi/2$, $n \in \mathbb{N}$, the $g$-factor vanishes as the structure is achiral; otherwise it is quite sensitive to the angle.
Figure~\ref{Figura_H}(b) demonstrates that simple chiral systems composed of only four dipoles may already exhibit substantial dissymmetry factors. 
Besides, this simple chiral system illustrates how resonant scattering may be explored to control the chiroptical response. This can be appreciated in Fig.~\ref{Figura_H}(c) where the sign of the $g$-factor changes with the detuning $\Delta$, which can be achieved by varying the driving frequency by approximately $\Gamma$.
Considering that atomic transitions can be very narrow, particularly for clock transitions, one obtains a structure in which chiroptical properties can be finely tuned, either with the driving frequency or, equivalently, with shifts imposed through external fields such as the Zeeman or Stark shifts. 

Finally, we notice that the chiroptical response of the twisted H is strongest when its size is smaller than the wavelength, with the $g$-factor going to zero when the system size is larger than $\approx 0.8\lambda$, see Fig.~\ref{Figura_H}(d). This can be understood as a manifestation of the $1/(kr)^2$ and $1/(kr)^3$ near-field terms in the dipole-dipole interaction~\eqref{green_dyadic}, which couple the different polarization channels and are strongest at short scales $r<\lambda$. From a different perspective, the vanishing of the $g$-factor for structures larger than the wavelength corroborates the expectation that chiroptical effects in light scattering arise from multiple scattering. Specifically, a nonzero chiroptical response requires at least four scattering events, enabling the optical field to effectively probe the chirality of the system and thus necessitating a minimum of four scatterers~\cite{ pinheiro2002magnetochiral}. The fast fluctuations of the $g$-factor in Fig.~\ref{Figura_H}(d) are a result of the collective eigenmodes of the system, whose contributions interfere to produce the scattered field following external drive. However, it is difficult to establish a direct correspondence between individual collective modes and the observed variations of the $g$-factor because the collective modes are not spectrally well separated. As a result, the scattered field arises from the simultaneous coherent excitation of multiple modes. The fast oscillations observed in the $g$-factor therefore originate from the strong sensitivity of the relative phases of these mode contributions to the laser detuning.

\begin{widetext}
\begin{figure*}[t]
\centering
\includegraphics[width=18.0cm,height=3.7cm]{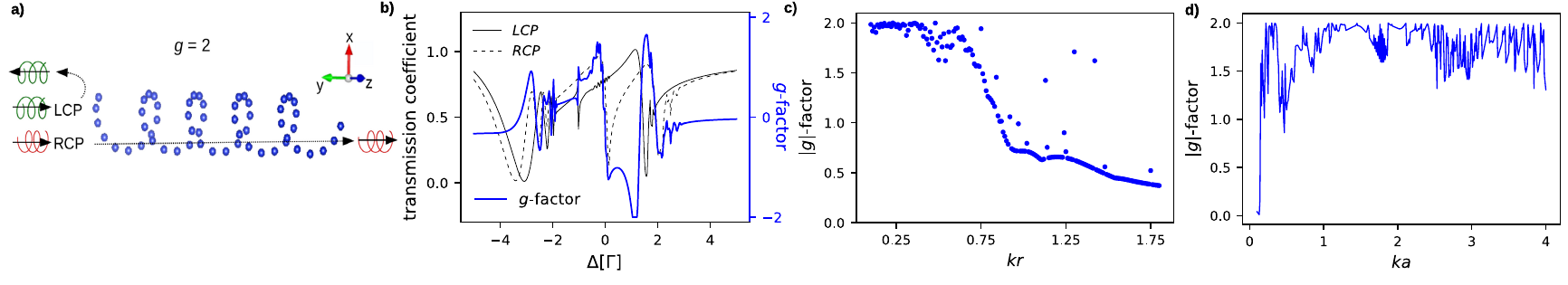}
\caption{(a) Representation of LCP and RCP propagating light through a helical atomic chain with right-handed helicity, all along the $z$-axis, for a $g$-factor with maximum value. The atomic positions of this structure are given by $(r\cos\theta,\:r\sin\theta,\:a\theta/2\pi)$, with $\theta =j 2\pi N_{rev}/N$, where $j = 1,...,N$ and $N_{rev}$ denotes the number of revolutions. (b) Transmission coefficients (black curves) for LCP (plain) and RCP (dashed) probe light, and the $g$-factor (blue curve), which quantifies the CD. Simulations realized for a pitch $a=1.75/k$, radius $r=0.5/k$, and $N=60$ atoms, with $3$ atoms per revolution. The nearest-neighbor separation for this configuration is approximately $0.87/k$. Inverted results are obtained for a left-handed helix. (c-d) $g$-factor as a function of the helix (c) radius and (d) pitch. While the CD is maximum for radius $r<0.6/k$, it maintains high values for a broad range of pitches. In order to obtain these results, for each value of $r$ and $a$ the highest $g$-factor is chosen considering a broad range of $\Delta$.}
\label{Figura_hélice}
\end{figure*}
\end{widetext}

{\em Atomic helical structures.} 
Motivated by the fact that the twisted H, the simplest chiral structure, already exhibits a substantial chiroptical response in light scattering from atoms, we now consider more complex configurations, such as atoms distributed along helices, the hallmark of chiral geometries. This atomic system, sketched in Fig.~\ref{Figura_hélice}(a), is known to exhibit strong intrinsic chirality~\cite{doi:10.1021/ja00278a029, electromagnetic_chir}. In Appendix I we monitor the effects of positional disorder in the chiroptical effects of the helix. Remarkably, large values of the $g$-factor are robust against fluctuations of the order of up to $1/k$. Similarly to the twisted H, the transmission of RCP and LCP probe beam is very sensitive to the frequency, but also quite different for each polarization, as it can be seen Fig.~\ref{Figura_hélice}(b). This results in a $g$-factor that reaches values close to $\pm2$, depending on the frequency. We thus conclude that the atomic helix exhibits strong chiroptical properties, which can be tuned by varying the atomic detuning, including the sign of the $g$-factor.

In order to achieve large chiroptical effects, the nearest-neighbor distance should be of the order of the optical wavelength. This condition ensures the occurrence of multiple scattering within the structure. The importance of considering subwavelengths atomic helices to enhance chiroptical effects is evidenced in Fig.~\ref{Figura_hélice}(c), where the $g$-factor remains maximal for a radius $r\lesssim 0.6/k$. Large chiroptical response in the subwavelength regime is uncommon for naturally occurring small scattering systems~\cite{2020LSA.....9..139M}, and typically require plasmonic particles arranged in typical chiral geometries such as helical assemblies of nanospheres~\cite{helical_nanospheres}, twisted nanorod dimers~\cite{chiral_dimers}, asymmetric pyramids of chiral nanoparticles~\cite{doi:10.1021/ja3066336}, or even disordered ensembles that also lack mirror symmetry~\cite{drachev2001large,pinheiro2017spontaneous}. Experimental implementations of subwavelength systems, in turn, have been realized successfully across various platforms, including cold atoms~\cite{PhysRevA.101.041603,PRXQuantum.6.010322}. Here the maximum $g$-factor exhibits a weak dependence on the helix pitch $a$, with values close to $2$ for a broad range of pitch values, see Fig.~\ref{Figura_hélice}(d). In the helical case the pitch is the parameter that governs the geometrical chiral properties~\cite{RevModPhys.71.1745}. Note that for the smallest values of $ka$, the system length for the $N = 60$ dipoles becomes smaller than the wavelength, and the chiroptical response vanishes for $ka< 0.05$; in this case the helix length becomes smaller than the wavelength.

While the above results correspond to an helix whose main axis coincides with the incident wavevector, the intrinsic chiroptical response of the structure is determined from the orientation-averaged $\left \langle CD \right \rangle_{\Omega}$, where $\left \langle \\\ \right \rangle_{\Omega}$ corresponds to an average over all possible orientations of the molecule~\cite{electromagnetic_chir}. The intrinsic chiroptical property $\left \langle CD \right \rangle_{\Omega}$ is of particular interest when the structures have an arbitrary orientation or no symmetry axis, such as  disordered systems. 
Here, this average is computed by integration over $4\pi$ solid-angle, so that the system is illuminated over all possible angles. For each structure, we choose the detuning which leads to the maximum chirality. For the twisted H structure, one obtains $|\left \langle CD \right \rangle_{\Omega}| = 0.02$, for $\Delta = -3.60\Gamma$, which indicates a weak but nonzero intrinsic chiroptical response. For the helical structure, $|\left \langle CD \right \rangle_{\Omega}| = 0.12$ for $N = 60$ and $\Delta = 1.62\Gamma$, and $|\left \langle CD \right \rangle_{\Omega}| = 0.54$ for $N = 120$ and $\Delta = -0.14\Gamma$, which show that a strong intrinsic chirality can be achieved for an elongated helical atomic structure.

\begin{figure}[t]
\centering
\includegraphics[width=\columnwidth]{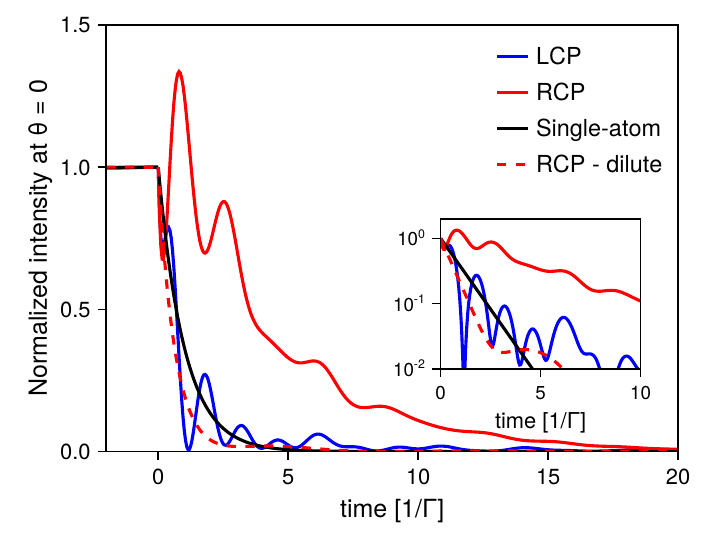}
\caption{Scattered intensity in a switch-off dynamics for a right-handed helix with $N = 60$, $r = 0.5/k$ and $a = 1.75/k$ for plain curves ($a = 5/k$ for the ``dilute'' case, see dashed curve), at resonance. A "chiral flash" and a subradiant dynamics can be observed for the RCP radiation in the regular (non-dilute) helix.}
\label{Figura_4}
\end{figure}

{\em Long-lived chiral emission.} The dipole-dipole coupling leads to not only energy shifts and dispersive properties, but also a strong modification of the emission lifetime~\cite{PhysRevLett.30.309, GROSS1982301, Inouye,2007NatPh...3..106S, de_Oliveira_2014,PhysRevLett.117.073002, jahnke,PhysRevA.95.043818, kuraptsev,wangZhen, GUERIN2023253}. Aiming to study the nature of the collective modes behind the chiroptical properties, we now investigate the radiative dynamics of the system after the laser is switched off, taking as initial condition a driven steady state. As shown in  Fig.~\ref{Figura_4}, the forward-scattered intensity, normalized by the steady state intensity, first exhibits a ``chiral flash'' for the RCP probe, with the dynamical intensity reaching a value larger than the steady-state one. Such flash is a transient phenomenon that occurs when a close-to-resonance laser illuminating an optically thick atomic cloud is abruptly switched off, leading to a coherent burst of light in the forward direction~\cite{PhysRevA.84.011401}. Since the helical structure under consideration is right-handed by construction, with $g = -0.25$, from Eq.~\eqref{eq_g_factor} we have that the LCP is better transmitted than the RCP beam ($T_{LCP} > T_{RCP}$), so that the flash occurs only for the less-transmitted polarization. This shows that coherent burst of light in the forward direction is also an optical manifestation of chirality in atomic systems, henceforth dubbed chiral flash.

Furthermore, at later times the scattered intensity exhibits a much slower decay than for isolated atom, see the inset in Fig.~\ref{Figura_4}. This subradiant dynamics has been reported in several platforms~\cite{PhysRevA.90.012511,Weiss_2018,PhysRevLett.128.203601,PhysRevA.106.053702, guerin2015}, with applications to storage and retrieval of light~\cite{PhysRevLett.117.243601,PhysRevLett.124.253603,Rui_2020,PhysRevX.11.021031}. This phenomenon is observed only in the RCP configuration. This can be interpreted as the fact that the optical depth in this channel is much larger than in the LCP one~\cite{guerin2015}, so the system exhibits a form of chiral subradiance. The long timescales are associated with the strong dipole-dipole interactions, since for a more dilute cloud, this subradiant dynamics is almost suppressed (see dashed curve in Fig.~\ref{Figura_4}). 
Note that subradiance is a slow radiative process distinct from multiple scattering~\cite{Weiss_2018,PhysRevA.104.023705}.

{\em Conclusions and outlook.} Detecting weak chiral signals remains a central challenge, as they often originate from the interference between electric and magnetic dipole moments, the latter typically being small. Here we show that purely electric-dipole systems can nevertheless exhibit pronounced chiroptical response, even the minimal chiral configuration composed of four atoms. Atomic assemblies of several tens of resonant emitters display strong intrinsic chirality and reach maximal $g$-factors. Importantly, the chiroptical response of these structures, including its sign, can be tuned either by varying the drive frequency or via externally induced level shifts. Moreover, the coexistence of superradiant and subradiant chiral modes provides a route toward controlling photon flow and storing optical information. Realizing such chiral architectures with cold atoms offers a promising pathway to combine strong chiral responses with nonlinear quantum-optical effects, for example using Rydberg-mediated interactions \cite{PhysRevLett.127.263602,Zhang2022photonphoton,Zhang:25}, enabling polarization- and photon-number–dependent scattering in free-space atomic arrays.

\begin{acknowledgments}
The authors acknowledge Dr. Noel A. Moreira for the fruitful discussions and support with the numerical Julia package. This study was financed, in part, by the Brazilian National Council for Scientific and Technological Development – CNPq, Grants No. 140248/2023-4 and 313632/2023-5. R.~B.~acknowledges the support from the S\~ao Paulo Research Foundation (FAPESP, Grants Nos. 2022/00209-6 and 2023/03300-7). F.A.P. acknowledges financial support from CAPES, CNPq, and FAPERJ.
\end{acknowledgments}





\bibliography{bibliography}

\newpage

\appendix

\section{Appendix I}
\label{apendice:appendix_1}

\begin{figure}[h]
\centering
\includegraphics[width=\columnwidth]{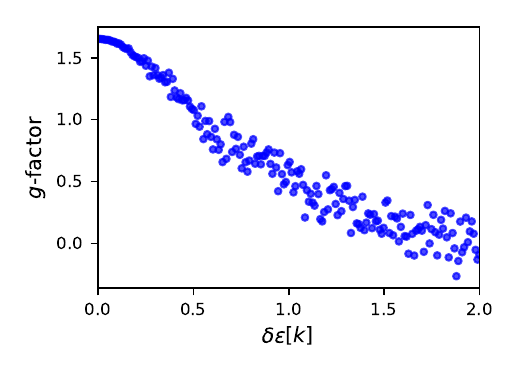}
\caption{$g$-factor for a helical structure with $N = 60$, $r = 0.5/k$ and $a = 1.75/k$. To each atom position, a random displacement $ \delta \vec{\epsilon}$ is added, leading to a progressive decrease of the chirality.}
\label{Fig_appendix}
\end{figure}

In Fig.~\ref{Fig_appendix} we present the $g$-factor as a function of the positional disorder on the atomic structure. The results were obtained by introducing a small variation $\delta\vec{\epsilon}$, of amplitude $\delta\epsilon/k$, to the position of each atom, such that $\vec{r}_{j} \rightarrow \vec{r}_{j} + \delta\vec{\epsilon}_{j}$. As it can be observed in Fig.~\ref{Fig_appendix}, remarkably, high values of the $g$-factor are preserved for values of this position disorder up to $\sim 0.5/k$.

\end{document}